\documentclass[aps,pre,twocolumn,showpacs,superscriptaddress]{revtex4}
\usepackage{epsfig}
\usepackage{times}
\begin{document}

\title{Degree mixing in multilayer networks impedes the evolution of cooperation}

\author{Zhen Wang}
\affiliation{Department of Physics, Hong Kong Baptist University, Kowloon Tong, Hong Kong}
\affiliation{Center for Nonlinear Studies and Beijing-Hong Kong-Singapore Joint Center for Nonlinear and Complex systems, Institute of Computational and Theoretical Studies, Hong Kong Baptist University, Kowloon Tong, Hong Kong}

\author{Lin Wang}
\affiliation{Centre for Chaos and Complex Networks, Department of Electronic Engineering,
City University of Hong Kong, Kowloon, Hong Kong}

\author{Matja{\v z} Perc}
\email{matjaz.perc@uni-mb.si}
\affiliation{Faculty of Natural Sciences and Mathematics, University of Maribor, Koro{\v s}ka cesta 160, SI-2000 Maribor, Slovenia}

\begin{abstract}
Traditionally, the evolution of cooperation has been studied on single, isolated networks. Yet a player, especially in human societies, will typically be a member of many different networks, and those networks will play a different role in the evolutionary process. Multilayer networks are therefore rapidly gaining on popularity as the more apt description of a networked society. With this motivation, we here consider $2$-layer scale-free networks with all possible combinations of degree mixing, wherein one network layer is used for the accumulation of payoffs and the other is used for strategy updating. We find that breaking the symmetry through assortative mixing in one layer and/or disassortative mixing in the other layer, as well as preserving the symmetry by means of assortative mixing in both layers, impedes the evolution of cooperation. We use degree-dependent distributions of strategies and cluster-size analysis to explain these results, which highlight the importance of hubs and the preservation of symmetry between multilayer networks for the successful resolution of social dilemmas.
\end{abstract}

\maketitle

\section{Introduction}
Evolutionary games on networks are the subject of intense recent exploration, as evidenced by current reviews that focus both on pairwise social dilemmas, such as the prisoner's dilemma and the snowdrift game \cite{szabo_pr07, roca_plr09, perc_bs10}, as well as on games that are governed by group interactions, such as the public goods game \cite{perc_jrsi13}. The subject has been made popular by the discovery that spatial structure can promote the evolution of cooperation \cite{nowak_n92b} through the mechanism that is now known as network reciprocity \cite{nowak_s06}. In essence, network reciprocity relies on the fact that cooperators do best if they are surrounded by other cooperators. If interactions amongst players are structured rather than well-mixed, the clustering of cooperators is more likely to be stable since defectors have limited opportunities for exploiting those that are located in the interior of cooperative clusters. Further promoting the potency of network reciprocity, which was initially studied predominantly on regular lattices \cite{nowak_ijbc94, szabo_pre98}, is heterogeneity of the interaction networks. Especially the fact that scale-free networks provide a unifying framework for the evolution of cooperation \cite{santos_prl05, santos_pnas06} has captured the attention of the physics community, as it became apparent that methods of statistical physics can be used successfully to study collective phenomena in social systems \cite{castellano_rmp09}, and in particular in evolutionary games \cite{perc_csf13}. Many works have since been devoted to the study of evolutionary games on small-world \cite{abramson_pre01, kim_bj_pre02, masuda_pla03, tomassini_pre06, vukov_pre06, fu_epjb07, vukov_pre08}, scale-free \cite{santos_pnas06, gomez-gardenes_prl07, rong_pre07, masuda_prsb07, tomassini_ijmpc07, szolnoki_pa08, assenza_pre08, santos_n08, pena_pre09, poncela_pre11, poncela_epl09, brede_epl11, tanimoto_pre12, pinheiro_pone12, simko_pone13}, coevolving \cite{ebel_pre02, zimmermann_pre04, szabo_pre04b, pacheco_prl06, santos_ploscb06, fu_pa07, tanimoto_pre07, fu_pre08b, fu_pre09}, as well as hierarchical \cite{vukov_pre05, lee_s_prl11} and bipartite \cite{gomez-gardenes_c11, gomez-gardenes_epl11} networks.

Although recent large-scale human experiments indicate that spatial reciprocity may be compromised or fail altogether \cite{gracia-lazaro_srep12, gracia-lazaro_pnas12}, there is still ample interest in understanding how and why networks influence the evolution of cooperation. The attention has recently been shifting towards the evolution of cooperation on interdependent and multilayer networks \cite{wang_z_epl12, gomez-gardenes_srep12, gomez-gardenes_pre12, wang_b_jsm12, wang_z_srep13, wang_z_srep13b, jiang_ll_srep13, szolnoki_njp13}. Indeed, several mechanisms have been discovered by means of which the interdependence between different networks or network layers may help to resolve social dilemmas. Examples include interdependent network reciprocity \cite{wang_z_srep13}, non-trivial organization of cooperators across the interdependent layers \cite{gomez-gardenes_srep12}, probabilistic interconnectedness \cite{wang_b_jsm12}, and information transmission \cite{szolnoki_njp13}. In addition to the evolution of cooperation on interdependent and multilayer networks, cascading failures \cite{buldyrev_n10, li_w_prl12, parshani_pnas11, brummitt_pnas12}, competitive percolation \cite{parshani_prl10, nagler_np11, cellai_pre13}, transport \cite{morris_prl12}, diffusion \cite{gomez_prl13}, neuronal synchronization \cite{sun_xj_chaos11}, epidemic spreading \cite{granell_prl13}, robustness against attack and assortativity \cite{huang_xq_pre11, zhou_d_pre12}, stability \cite{cozzo_pre13}, growth \cite{nicosia_prl13}, entropy and overlap \cite{bianconi_pre13}, as well as abrupt transition in the structural formation \cite{radicchi_np13}, have also been studied. Networks of networks have captivated the attention of large contingents of natural and social sciences \cite{gao_jx_np12, havlin_pst12, helbing_n13, csermely_tde13}, and a comprehensive review is already available that survey the rapidly increasing literature \cite{kivela_ax13}.

Here we wish to extend the scope of evolutionary games on multilayer networks by studying the impact of degree mixing on $2$-layer scale-free networks. One layer thereby serves as the interaction network where players accumulate their payoffs, while the other layer serves as the updating network where players change their strategies. This setup takes into account the fact that especially in human societies each individual is member in many different networks, and those networks typically play very different roles. The distinction of interaction and updating networks is akin to previous works that studied the impact of symmetry breaking between interaction and replacement in evolutionary games on graphs \cite{ohtsuki_prl07, ohtsuki_jtb07b}. There it has been concluded that it is always harder for cooperators to evolve whenever the two graphs do not coincide, and our current results will support such a conclusion further. In layered networks, for interdependent network reciprocity to work \cite{wang_z_srep13}, the simultaneous formation of correlated cooperative clusters on both networks is crucial, which however is disturbed if the networks do not overlap or are insufficiently interconnected \cite{wang_z_srep13b}. In terms of degree mixing, we follow the explorations by Rong et al. \cite{rong_pre07}, who studied the role of assortative and disassortative mixing on the evolution of cooperation on isolated scale-free networks. The study concluded that assortative mixing by degree promotes defection because of the increase of the interconnectedness of hubs, while disassortative mixing may prevent the extinction of cooperators because isolated hubs act as a safe refuge against invading defectors. As we will show when presenting the main results, our study interpolates between the previous findings concerning the symmetry breaking between interaction and replacement, and the role of degree mixing on isolated scale-free networks.

The remainder of this paper is organized as follows. First, we describe the mathematical model, in particular the procedure for the construction of multilayer scale-free networks with assortative and disassortative mixing, as well as the definition and the simulation protocol of social dilemmas. Next we present the main results, whereas lastly we summarize and discuss their implications.

\section{Mathematical model}

\begin{figure}
\centerline{\epsfig{file=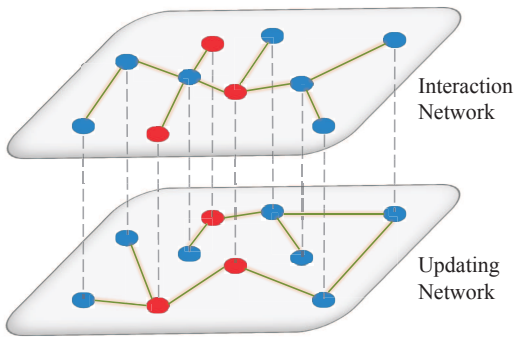,width=7cm}}
\caption{(Color online) Schematic presentation of a $2$-layer system, consisting of the interaction and the updating network. In the interaction network players obtain their payoffs, while in the updating network they look for neighbors to change their strategies. In the depicted example the two layers differ in their structure. We achieve this by applying the Xulvi-Brunet-Sokolov algorithm \cite{xulvi-brunet_pre04} with different assortative/disassortative coefficient $\mathcal{A}$ (denoted by $\mathcal{A}_I$ and $\mathcal{A}_U$) on each individual layer. If the applied value of $\mathcal{A}$ is the same for both layers ($\mathcal{A}_I = \mathcal{A}_U$), the symmetry between the interaction and the updating network is preserved, and the setup becomes identical to a single scale-free network subject to degree mixing. We use the ``multilayer network'' terminology to emphasize the important conceptual link between this theoretical framework and previous works, where edge colored networks or ``breaking the symmetry between interaction and replacement'' \cite{ohtsuki_prl07, ohtsuki_jtb07b} have also been used.}
\label{scheme}
\end{figure}

The $2$-layer scale-free networks are constructed as follows. We first use the algorithm of Barab{\'a}si and Albert \cite{barabasi_s99} to construct a neutrally degree-mixing scale-free network with an average degree $<k>=4$. Importantly, the algorithm may introduce spurious correlations that violate neutral mixing if the thermodynamic limit is not reached. To ensure a proper baseline setup without degree-mixing, we therefore use the Xulvi-Brunet-Sokolov algorithm \cite{xulvi-brunet_pre04} to remove such correlations. Alternatively, the configurational model by Molloy and Reed \cite{molloy_rsa95} could also have been used to generate scale-free networks without spurious correlations directly, thus alleviating the need for further adjustments. Subsequently, again using the Xulvi-Brunet-Sokolov algorithm \cite{xulvi-brunet_pre04}, we produce a series of degree-mixing networks, where $\mathcal{A}$ is the assortative/disassortative coefficient. Assortative mixing (i.e., $\mathcal{A}>0$) introduces the tendency for nodes with similar degree to become directly connected, while disassortative mixing (i.e., $\mathcal{A}<0$) introduces the tendency for nodes with similar degree to become disconnected. Since the coefficient $\mathcal{A}$ of most empirical networks falls into the interval $[-0.3, 0.3]$ \cite{rong_pre07}, we focus on this range when presenting the main results in Section III. If both layers are characterized by the same value of $\mathcal{A}$, then they are completely identical.

\begin{figure}
\centerline{\epsfig{file=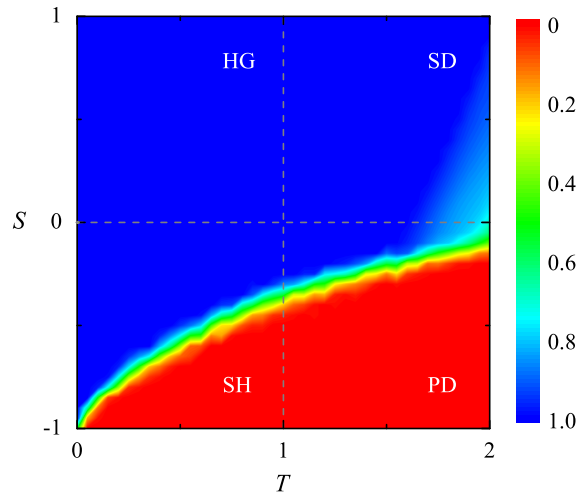,width=7.5cm}}
\caption{(Color online) Color map encoding the fraction of cooperators $\rho_C$ on the $T-S$ parameter plane, as obtained on an isolated scale-free network without degree mixing. Within our model this traditional setup is recovered for $\mathcal{A}_I=\mathcal{A}_U=0$. In agreement with known results (see \cite{santos_pnas06}), the scale-free network is able to sustain cooperation even for unfavorable combinations of $T>0$ and $S<0$. We also delineate different quadrants, which represent different social dilemmas. The $T<1$ and $S>1$ quadrant marks the harmony game (HG), which however does not constitute a social dilemma. For the interpretation of color with regards to $\rho_C$ see the vertical color bar on the right. Due to the fine mesh, the interpolation of color (or lack thereof) has no visible effect on the presentation of the results.}
\label{original}
\end{figure}

Each player $x$ is initially designated either as a cooperator (strategy $s_x=C$) or defector (strategy $s_x=D$) with equal probability, and it is simultaneously located on both networks. One is the interaction network, where players play the evolutionary game (to be introduced below) to obtain their payoffs, while the other is the updating network, where players seek for neighbors to potentially update their strategy. To take into account the fact that the interaction and the updating network may differ, we distinguish the values of $\mathcal{A}$ for both networks. We use $\mathcal{A}_I$ and $\mathcal{A}_U$ to denote the assortive/disassortive coefficient of the interaction and the updating network, respectively. The setup of the $2$-layer network where $\mathcal{A}_I \neq \mathcal{A}_U$ is depicted schematically in Fig.~\ref{scheme}.

The evolutionary social dilemmas are defined as follows. Mutual cooperation yields the reward $R$, mutual defection leads to punishment $P$, and the mixed choice gives the cooperator the sucker's payoff $S$ and the defector the temptation $T$. Within this traditional setup we have the prisoner's dilemma (PD) game if $T>R>P>S$, the snowdrift game (SD) if $T>R>S>P$, and the stag-hunt (SH) game if $R>T>P>S$, thus covering all three major social dilemma types where players can choose between cooperation and defection. Following common practice \cite{szabo_pr07}, we set $R = 1$ and $P=0$, thus leaving the remaining two payoffs to occupy $-1 \leq S \leq 1$ and $0 \leq T \leq 2$, as depicted schematically in Fig.~\ref{original}.

We simulate the evolutionary process in accordance with the standard Monte Carlo simulation procedure comprising the following elementary steps. First, a randomly selected player $x$ acquires its payoff $P_x$ by playing the game with all its neighbors on the interaction network. Next, player $x$ randomly chooses one neighbor $y$ on the updating network, who then also acquires its payoff $P_y$ on the interaction network in the same way as previously player $x$. Lastly, player $x$ adopts the strategy $s_y$ from player $y$ with a probability determined by the Fermi function \cite{blume_l_geb93, szabo_pre98}
\begin{equation}
W(s_y \to s_x)=\frac{1}{1+\exp[(P_x-P_y)/K]},
\end{equation}
where $K=0.1$ quantifies the uncertainty related to the strategy adoption process. The selected value of $K$ is a traditional and frequently employed choice that does not qualitatively affect the evolutionary outcomes, as shown in many preceding works and reviewed comprehensively in \cite{szabo_pr07}. Each full Monte Carlo step (MCS) gives a chance for every player to change its strategy once on average. The baseline outcome is depicted in Fig.~\ref{original}, which is recovered if we set $\mathcal{A}_I=\mathcal{A}_U=0$ when constructing the $2$-layer interaction network. For the presentation of the results, we employ a color mapping of the fraction of cooperators $\rho_C$ on the $T-S$ plane, as used recently in \cite{roca_plr09, cardillo_njp10, buesser_pre12b}, whereby the employed mesh encompasses $81 \times 81$ parameter combinations. In what follows, we will present the main results for all possible combinations of $\mathcal{A}_I$ and $\mathcal{A}_U$, first for the symmetry preserving $\mathcal{A}_I=\mathcal{A}_U$ case, and subsequently for the symmetry breaking $\mathcal{A}_I \neq \mathcal{A}_U$ case. The simulation results are typically obtained for scale-free networks with $10^4$ nodes, and the stationary fraction of cooperators $\rho_C$ is determined as the average within the last $10^4$ out of the total $10^5$ MCS. Naturally, close to the transition points to pure $C$ and $D$ phases the system size needs to be increased to avoid an accidental extinction of the subordinate strategy, and the simulations times must be accordingly longer. We have taken this into account by using larger system size where needed, and the simulations were performed until the stationary state was reached. In general, the equilibrium (or the stationary state) is reached when the average of the cooperation level becomes time-independent. Moreover, since the structure of assortative/disassortative networks and random distributions of strategies may introduce additional uncertainty, the final results are averaged over up to $100$ independent runs for each set of parameter values in order to assure suitable accuracy.

\section{Results}

\subsection{Symmetry preservation}

\begin{figure}
\centerline{\epsfig{file=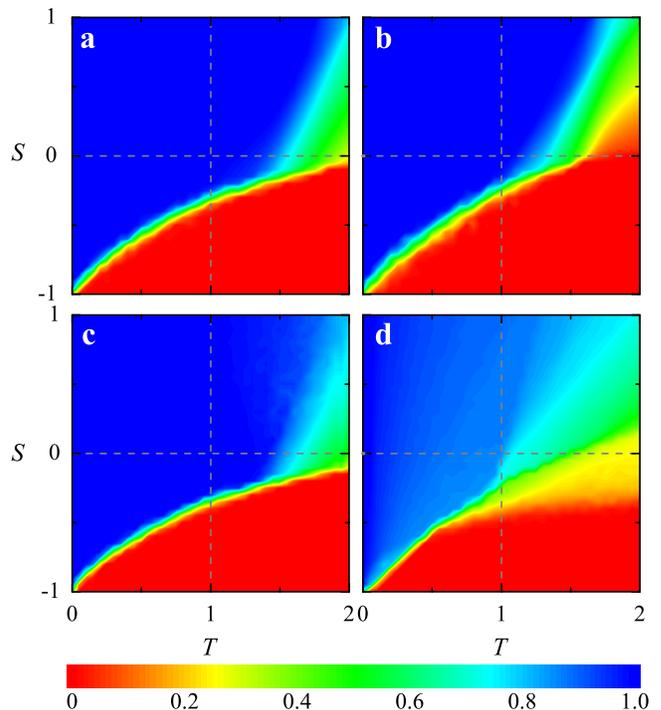,width=8.5cm}}
\caption{(Color online) Color map encoding the fraction of cooperators $\rho_C$ on the $T-S$ parameter plane, as obtained in the symmetry preserving $\mathcal{A}_I=\mathcal{A}_U$ case. Top two panel: assortative mixing is applied on both networks. Bottom two panels: disassortative mixing is applied on both networks. Comparing with the baseline (see Fig.~\ref{original}), the evolution of cooperation is substantially inhibited. An exception is strong disassortative mixing (panel \textbf{d}), where cooperators are able to survive at slightly harsher conditions than for the baseline case. From panel \textbf{a} to \textbf{d}, the values of the assortative/disassortative coefficient $\mathcal{A}_I$ and $\mathcal{A}_U$ are $0.1$, $0.3$, $-0.1$ and $-0.3$, respectively. For the interpretation of color with regards to $\rho_C$ see the horizontal color bar at the bottom. Like in Fig.~\ref{original}, the interpolation of color has no visible effect on the presentation of the results.}
\label{symmetry}
\end{figure}

To begin with, we present the results obtained for the symmetric degree mixing of both scale-free network layers. In this case, the assortative/disassortative coefficient is thus identical for both networks ($\mathcal{A}_I=\mathcal{A}_U$), which returns our setup to the already studied single degree-mixing network \cite{rong_pre07}. Figure~\ref{symmetry} shows the color map encoding the fraction of cooperation $\rho_C$ on $T-S$ parameter plane for four different values of $\mathcal{A}_I=\mathcal{A}_U$. For the assortative mixing (top two panels), it is clear that, compared with the baseline performance depicted in Fig. ~\ref{original}, cooperative behavior is restrained and the dominance space of full cooperation shrinks. In particular, the larger the value of $\mathcal{A}_I$ ($\mathcal{A}_U$), the more obvious the trait of inhibition. This is caused by the changes in the topology that are due to assortative mixing. In particular, large-degree hubs tend to interconnect with each other, which destroys the sustainability of cooperative clusters and promotes the invasion of defectors. If disassortative mixing is applied (bottom two panels), we can see that, in the majority of the parameter space, the level of cooperation is again lower than what we have observed in the absence of mixing. However, under harsh conditions, where the temptation to defect is large and the sucker's payoff is negative (the PD quadrant), we find that cooperation is a slightly more persistent. Due to the absence of direct links between large-degree hubs, the isolated cooperator nodes can successfully resist the invasion of defectors and hold their initial strategy, even at conditions where the neutrally mixing scale-network fails to sustain cooperative behavior. Along this line, if the disassortative mixing would be even stronger, we can predict that this phenomenon may become more noteworthy and extent across a larger are of the $T-S$ plane. These results are in agreement with \cite{rong_pre07}, and they provide an informed entry into the study of asymmetric mixing ($\mathcal{A}_I \neq \mathcal{A}_U$), which we consider next.

\subsection{Symmetry breaking}

\begin{figure}
\centerline{\epsfig{file=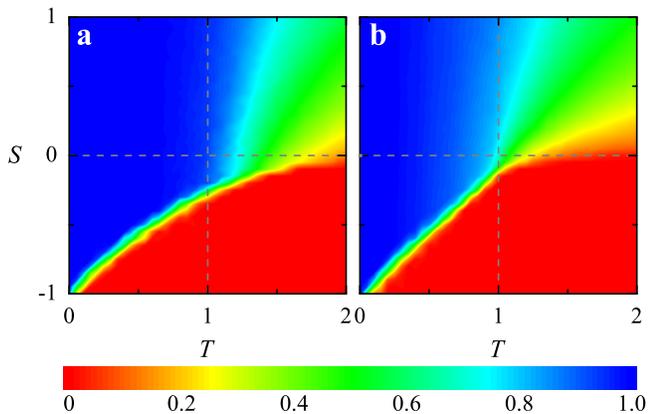,width=8.5cm}}
\caption{(Color online) Color map encoding the fraction of cooperators $\rho_C$ on the $T-S$ parameter plane, as obtained for the symmetry breaking assortative mixing of the interaction network ($\mathcal{A}_I>0$) and the disassortative mixing of the updating network ($\mathcal{A}_U<0$). The presented results indicate that the evolution of cooperation is impaired more severely if the mixing is stronger. Parameter values are $\mathcal{A}_I=0.1$, $\mathcal{A}_U=-0.1$ in panel \textbf{a} and $\mathcal{A}_I=0.2$, $\mathcal{A}_U=-0.2$ in panel \textbf{b}, respectively. Interpretation and interpolation of color is the same as in Fig.~\ref{symmetry}.}
\label{case2}
\end{figure}

In this section, we proceed with exploring the evolution of cooperation under the different asymmetric degree mixing options of the interaction and the updating network layer. For completeness, we consider all possible combinations of assortative and disassortative mixing on both layers.

Results presented in Fig.~\ref{case2} show the outcome obtained when the interaction network is subject to assortative mixing ($\mathcal{A}_I>0$) while the updating layer is subject to disassortative mixing ($\mathcal{A}_U<0$). Compared to the baseline (see Fig.~\ref{original}), this combination fails to promote cooperation, and indeed in the majority of the parameters space the evolution of cooperation is inhibited, especially in the PD and SD quadrant. Even at small temptations to defect the complete dominance of cooperators is no longer achievable, and as $T$ increases further, cooperative behavior fails faster than in the absence of mixing. The threshold marking extinction of cooperators increases as well. Although the assortative mixing of the interaction network may bestow higher payoffs to cooperative hubs, this advantage fails to manifest because the updating network is disassortative. Defectors, even if their degree is low, have access to the hubs and can thus invade effectively.

\begin{figure}
\centerline{\epsfig{file=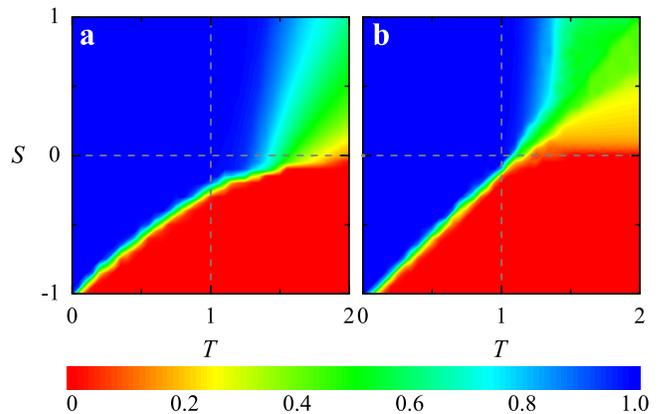,width=8.5cm}}
\caption{(Color online) Color map encoding the fraction of cooperators $\rho_C$ on the $T-S$ parameter plane, as obtained for the symmetry breaking disassortative mixing of the interaction network ($\mathcal{A}_I<0$) and the assortative mixing of the updating network ($\mathcal{A}_U>0$). As in the opposite case (Fig.~\ref{case2}), the evolution of cooperation is impaired compared to the baseline, thus supporting the disruptive effect of symmetry breaking. Parameter values are $\mathcal{A}_I=-0.1$, $\mathcal{A}_U=0.1$ in panel \textbf{a} and $\mathcal{A}_I=-0.2$, $\mathcal{A}_U=0.2$ in panel \textbf{b}, respectively. Interpretation and interpolation of color is the same as in Fig.~\ref{symmetry}.}
\label{case3}
\end{figure}

Naturally, it is also of interest to investigate the evolution of cooperation in the opposite case, namely if the interaction layer is subject to disassortative mixing ($\mathcal{A}_I<0$) while the updating layer is subject to assortative mixing ($\mathcal{A}_U>0$). We present in Fig.~\ref{case3} the results obtained for this particular mixing combination, and we use the same absolute values for $\mathcal{A}_I$ and $\mathcal{A}_U$ as in Fig.~\ref{case2} for an easier direct comparison. The goal is to test whether here the evolution of cooperation is also impaired. Compared with the results presented in Fig.~\ref{original}, the conclusion is again that, although the coexistence of cooperators and defectors is slightly extended on the $T-S$ plane, the negative effect of symmetry breaking nullifies the advantage of cooperative clusters and impairs the evolution of cooperation. Here interconnected hubs might reinforce their cooperation in the updating network, but the interaction network, where hubs are disconnected, fails to support this with appropriately high payoffs. Accordingly, we conclude that if assortative and disassortative networks make up different layers of the complex system, the multilayer combination does not promote the evolution of cooperation, regardless of the type of combination.

\begin{figure}
\centerline{\epsfig{file=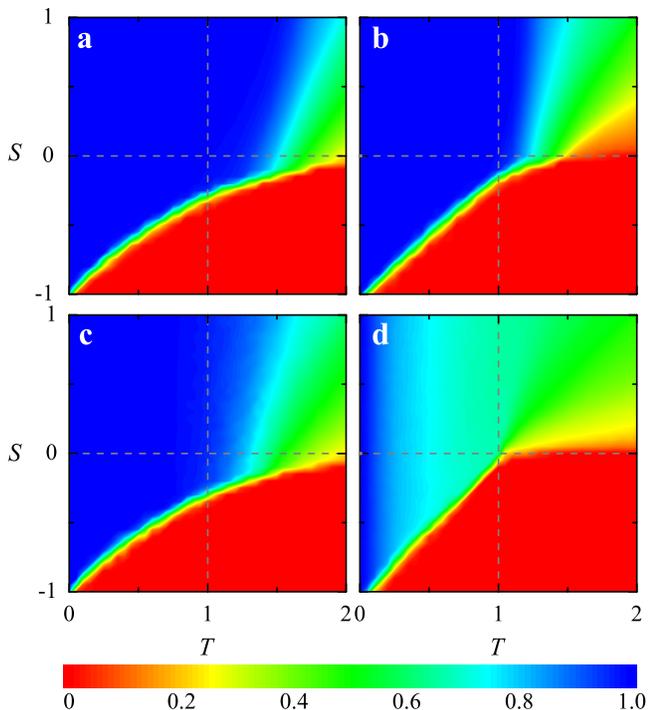,width=8.5cm}}
\caption{(Color online) Color map encoding the fraction of cooperators $\rho_C$ on the $T-S$ parameter plane, as obtained for the symmetry breaking neutral mixing of the interaction network ($\mathcal{A}_I=0$), the assortative mixing of the updating network ($\mathcal{A}_U>0$) (top two panels), and the disassortative mixing of the updating network ($\mathcal{A}_U<0$) (bottom two panels). If only the updating network layer is subject to degree mixing, the evolution of cooperation is impaired as well, and this regardless of the type of mixing. Parameter values for panels \textbf{a} to \textbf{d} are $\mathcal{A}_U$ are $0.1$, $0.3$, $-0.1$ and $-0.3$, respectively. Interpretation and interpolation of color is the same as in Fig.~\ref{symmetry}.}
\label{case4}
\end{figure}

Aside from the two options considered in Figs.~\ref{case2} and \ref{case3}, however, there are still further asymmetric setups that must be explored. In particular, we have to consider the possibility that only a single layer is subject to mixing, while the other remains degree-neutral. With this option in mind, we first show in Fig.~\ref{case4} how cooperators fare if the interaction network remains neutral ($\mathcal{A}_I=0$), while the updating layer is subject either to assortative mixing (top two panels) or disassortative mixing (bottom two panels). For the assortative updating network, it is clear that network reciprocity fails to protect cooperators against the exploitation by defectors sooner than for the baseline case. Both threshold values, marking the extinction of cooperators and defectors, increase with the increase of the assortative coefficient. As evidenced in the upper right panel, cooperators can only hold their undisputed dominance within a limited region (focusing on the most demanding PD and the SD quadrant) and go extinct even at moderate temptations to defect, especially in the snowdrift quadrant. This may be related to the preference for a checkerboard structure (on regular lattices), where mixed strategies warrant the highest payoff in the snowdrift game. Interestingly, the situation is even worse if the mixing of the updating layer is disassortative. Here (in the bottom two layers), not only does the overall cooperation level decreases, but the dominance of cooperators becomes impossible even for small temptations to defect as well. These results indicate that, even if applied to a single layer, degree mixing does not promote cooperation, especially not if the symmetry between different layers is broken.

\begin{figure}
\centerline{\epsfig{file=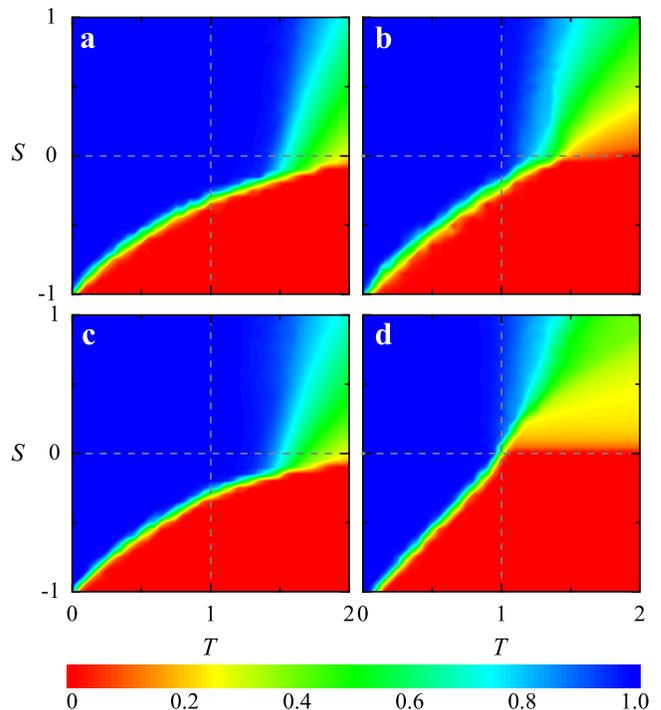,width=8.5cm}}
\caption{(Color online) Color map encoding the fraction of cooperators $\rho_C$ on the $T-S$ parameter plane, as obtained for the symmetry breaking neutral mixing of the updating network ($\mathcal{A}_U=0$), the assortative mixing of the interaction network ($\mathcal{A}_I>0$) (top two panels), and the disassortative mixing of the interaction network ($\mathcal{A}_I<0$) (bottom two panels). As in Fig.~\ref{case4}, if only the interaction network layer is subject to degree mixing, the evolution of cooperation is also impaired. Parameter values for panels \textbf{a} to \textbf{d} are $\mathcal{A}_I$ are $0.1$, $0.3$, $-0.1$ and $-0.3$, respectively. Interpretation and interpolation of color is the same as in Fig.~\ref{symmetry}.}
\label{case5}
\end{figure}

To conclude this section, we present results obtained with the last outstanding option, which is that the updating network remains neutrally mixing while the interaction network is subject to either assortative or disassortative mixing. Figure~\ref{case5} shows the results, which agree with those presented for all the other options, and which thus further support the conclusion that degree mixing in multilayer networks impedes the evolution of cooperation. As can be observed, and in agreement with the trends outlined thus far, the larger the absolute values of the coefficient $\mathcal{A}_I$, i.e., the stronger the mixing and the larger the symmetry breaking between the interaction and the updating network, the lesser the evolutionary success of the cooperative behavior across the $T-S$ plane. The failure of cooperation in the presence of degree mixing on multilayer networks can be understood and corroborated with an analysis of the dynamical organization of cooperative clusters \cite{gomez-gardenes_prl07}, which we will attend to in the next section.

\subsection{Analysis of cooperator clusters}

\begin{figure}
\centerline{\epsfig{file=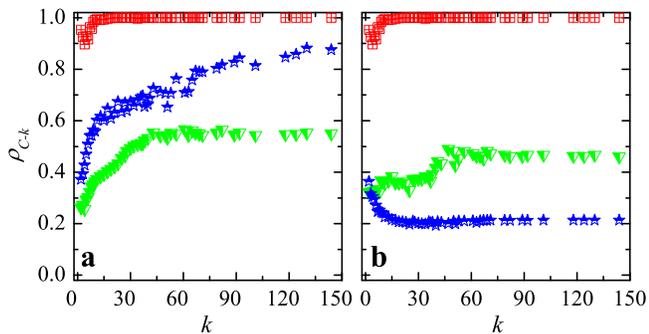,width=8.5cm}}
\caption{(Color online) Distributions of cooperators $\rho_{C-k}$ in dependence on the node degree $k$ on multilayer networks subject to different degree-mixing combinations. In both panels, red squares denote the distribution of cooperators for the baseline (i.e., $\mathcal{A}_I=\mathcal{A}_U=0$). It can be seen that cooperators occupy the hub nodes, which in turn attract a great number of followers to form giant cooperative clusters. Panel \textbf{a} shows the distribution for the symmetric case, where the values of assortative/disassortative coefficients are $\mathcal{A}_I=\mathcal{A}_U=-0.3$ (blue stars) and $\mathcal{A}_I=\mathcal{A}_U=0.3$ (green triangles). Panel \textbf{b} displays the distribution for a typical asymmetric case, where the values of assortative/disassortative coefficient are $\mathcal{A}_I=-0.2$, $\mathcal{A}_U=0.2$ (blue stars) and $\mathcal{A}_I=0.2$, $\mathcal{A}_U=-0.2$ (green triangles). Irrespective of whether the symmetry is preserved or broken, degree mixing in the studied multilayer networks decreases the ability of cooperators to hold onto the hubs of the network. All the potential followers therefore become more susceptible to defector invasions, and consequently the overall density of cooperators decreases. Presented results were obtained for $T=1.9$ and $S=0.5$, but remain qualitatively the same also for other $T-S$ combinations. The error bars in all panels are comparable to two times the symbol size.}
\label{degree}
\end{figure}

\begin{figure}
\centerline{\epsfig{file=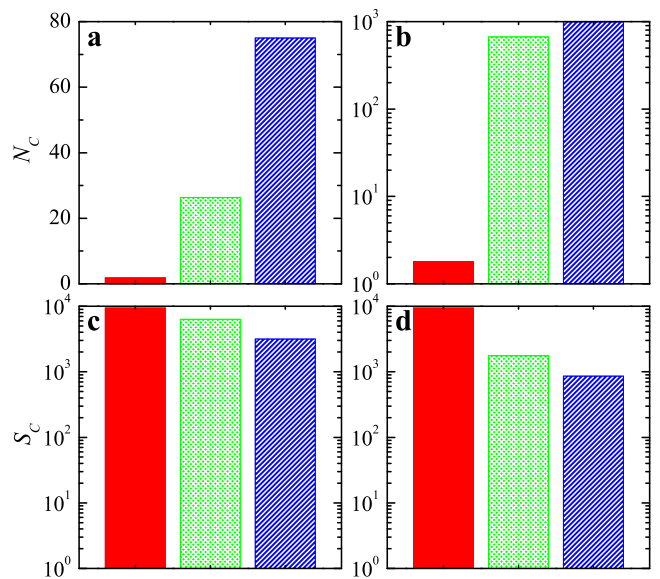,width=8.5cm}}
\caption{(Color online) Number of cooperative clusters $N_C$ (panels \textbf{a} and \textbf{b}) and the size of the largest cooperative cluster $S_C$ (panels \textbf{c} and \textbf{d}), as obtained for different degree-mixing combinations. Panels \textbf{a} and \textbf{c} show the results for symmetric mixing, where the middle column (green dotted fill) corresponds to $\mathcal{A}_I=\mathcal{A}_U=-0.3$ (disassortative mixing), while the right column (blue dashed fill) corresponds to $\mathcal{A}_I=\mathcal{A}_U=0.3$ (assortative mixing). Panels \textbf{b} and \textbf{d} show the results for asymmetric mixing, where the middle column (green dotted fill) corresponds to $\mathcal{A}_I=0.2$, $\mathcal{A}_U=-0.2$, while the right column (blue dashed fill) corresponds to $\mathcal{A}_I=-0.2$, $\mathcal{A}_U=0.2$. In all panels, the left column (red solid fill) depicts the result for the baseline case ($\mathcal{A}_I=\mathcal{A}_U=0$), where the small values of $N_C$ indicate that cooperators form giant clusters (typically there are only one or two clusters per network) to protect themselves against the invasion of defectors. When mixing is introduced, however, it can be observed that the cooperative clusters become more common and smaller compared to the baseline case. If in addition to mixing also the symmetry between the interaction and updating layers is broken, the disintegration of the large cooperative cluster on many small cooperative clusters is even more dramatic (note the log scale vertically). Presented results were obtained for $T=1.9$ and $S=0.5$, but remain qualitatively the same also for other $T-S$ combinations. The error bars on the columns are too small for visual display.}
\label{cluster}
\end{figure}

Although the simulation results yield a conclusive formulation of the impact of degree mixing on the evolution of cooperation in multilayer networks, one may still be curious as to why that is the case. While we have outlined some heuristic arguments when presenting the results, a more quantitative insight can be obtained by studying degree-dependent distributions of strategies and performing a cluster-size analysis. Figure~\ref{degree} features the distribution of cooperators in dependence on the degree of nodes for different combinations of mixing. For the baseline case (neutral mixing with symmetry breaking), we recover well-known results \cite{santos_prl05, gomez-gardenes_prl07}, according to which cooperators generally occupy the hubs of the network (although cooperation is possible in the presence of defector hubs too \cite{poncela_epl09}). Results in Fig.~\ref{cluster} further complement this picture by demonstrating that cooperators form a giant cooperative cluster and thereby make the most out of network reciprocity \cite{nowak_n92b}. Degree mixing, however, distorts this setup. Clusters disintegrate, and they become smaller. Cooperators are no longer able to hold onto the hubs, and the defectors have an easier time invading the smaller and more vulnerable cooperative domains. The stronger the mixing and the stronger the symmetry breaking between the interaction and updating network, the more complete and dramatic the disintegration becomes. The organizational efficiency of cooperators decays significantly, and it is interesting to discover that none of the possible combinations of mixing in multilayer networks (except for the symmetry-preserving disassortative mixing under adverse conditions) is able to improve the baseline support that is awarded to cooperators on isolated neutrally mixing scale-free networks.

\section{Discussion}
We have studied the evolution of cooperation on multilayer scale-free networks subject to assortative and disassortative mixing. We have considered all three main social dilemma types, and all possible combinations of assortativity and disassortativity. We have shown that, only if the symmetry between the interaction and the updating network is preserved, does the isolation of the hubs that is due to disassortative mixing sustain cooperation at harsher conditions than an isolated neutrally mixing scale-free network. If the symmetry between the interaction and the updating network is preserved but the mixing is assortative, the evolution of cooperation is impaired because the increasing interconnectedness of hubs favors the invasion of defectors. These results agree with the preceding work performed on an isolated scale-free network that is subject to degree mixing \cite{rong_pre07}. On the other hand, if the symmetry between the interaction and the updating network is broken, then any combination of assortative and disassortative mixing, regardless to which layer it applies to, impairs the evolution of cooperation. This conclusion applies to all social dilemmas, although most affected by degree mixing and the symmetry breaking are the snowdrift and the prisoner's dilemma quadrant. These results, on the other hand, agree with the exploration of symmetry breaking between interaction and replacement, as studied by Ohtsuki et al. \cite{ohtsuki_prl07}. The degree-dependent distributions of strategies reveal that degree mixing on multilayer networks hinders the ability of cooperators to persistently occupy the hubs, which in turn impairs their ability to make use of the enhanced network reciprocity that could be warranted by the degree heterogeneity of the interaction and the updating network \cite{santos_prl05}. This conclusion is strengthened and quantitatively corroborated by the investigation of the number and the size of cooperative clusters. The latter become more common and smaller as soon as the two layers loose symmetry, and regardless of how the symmetry breaking is introduced, i.e., whether the interaction network is subject to assortative mixing and the updating layer is subject to disassortative mixing, or vice versa, or if only a single network layer is subject to either type of mixing while the other layer remains degree neutral. These results thus interpolate between the impact of degree mixing on isolated networks, and the impact of symmetry breaking between interaction and replacement, and by doing so they provide further insight that fosters our understanding of the evolution of cooperation on multilayer networks. A specific point that merits further research is the consideration of different time scales related to the interaction and replacement, as already noted in \cite{ohtsuki_prl07} and studied prominently in \cite{roca_prl06}. Along with related recent discoveries \cite{wang_z_epl12, gomez-gardenes_srep12, gomez-gardenes_pre12, wang_b_jsm12, wang_z_srep13, wang_z_srep13b, jiang_ll_srep13, szolnoki_njp13}, as well as many other phenomena that are currently investigated in the realm of multilayer networks in statistical physics \cite{kivela_ax13}, we hope that our study will help contribute to the continued vibrancy of this field of research.

\begin{acknowledgments}
This research was supported by the Slovenian Research Agency (grant J1-4055). Helpful discussions with Yan Zhang and Changsong Zhou are gratefully acknowledged and much appreciated as well.
\end{acknowledgments}


\begin{thebibliography}{87}
\expandafter\ifx\csname natexlab\endcsname\relax\def\natexlab#1{#1}\fi
\expandafter\ifx\csname bibnamefont\endcsname\relax
  \def\bibnamefont#1{#1}\fi
\expandafter\ifx\csname bibfnamefont\endcsname\relax
  \def\bibfnamefont#1{#1}\fi
\expandafter\ifx\csname citenamefont\endcsname\relax
  \def\citenamefont#1{#1}\fi
\expandafter\ifx\csname url\endcsname\relax
  \def\url#1{\texttt{#1}}\fi
\expandafter\ifx\csname urlprefix\endcsname\relax\def\urlprefix{URL }\fi
\providecommand{\bibinfo}[2]{#2}
\providecommand{\eprint}[2][]{\url{#2}}

\bibitem[{\citenamefont{Szab{\'o} and F{\'a}th}(2007)}]{szabo_pr07}
\bibinfo{author}{\bibfnamefont{G.}~\bibnamefont{Szab{\'o}}} \bibnamefont{and}
  \bibinfo{author}{\bibfnamefont{G.}~\bibnamefont{F{\'a}th}},
  \bibinfo{journal}{Phys. Rep.} \textbf{\bibinfo{volume}{446}},
  \bibinfo{pages}{97} (\bibinfo{year}{2007}).

\bibitem[{\citenamefont{Roca et~al.}(2009)\citenamefont{Roca, Cuesta, and
  S{\'a}nchez}}]{roca_plr09}
\bibinfo{author}{\bibfnamefont{C.~P.} \bibnamefont{Roca}},
  \bibinfo{author}{\bibfnamefont{J.~A.} \bibnamefont{Cuesta}},
  \bibnamefont{and}
  \bibinfo{author}{\bibfnamefont{A.}~\bibnamefont{S{\'a}nchez}},
  \bibinfo{journal}{Phys. Life Rev.} \textbf{\bibinfo{volume}{6}},
  \bibinfo{pages}{208} (\bibinfo{year}{2009}).

\bibitem[{\citenamefont{Perc and Szolnoki}(2010)}]{perc_bs10}
\bibinfo{author}{\bibfnamefont{M.}~\bibnamefont{Perc}} \bibnamefont{and}
  \bibinfo{author}{\bibfnamefont{A.}~\bibnamefont{Szolnoki}},
  \bibinfo{journal}{BioSystems} \textbf{\bibinfo{volume}{99}},
  \bibinfo{pages}{109} (\bibinfo{year}{2010}).

\bibitem[{\citenamefont{Perc et~al.}(2013)\citenamefont{Perc,
  G{\'o}mez-Garde{\~n}es, Szolnoki, and Flor{\'{\i}a and Y.
  Moreno}}}]{perc_jrsi13}
\bibinfo{author}{\bibfnamefont{M.}~\bibnamefont{Perc}},
  \bibinfo{author}{\bibfnamefont{J.}~\bibnamefont{G{\'o}mez-Garde{\~n}es}},
  \bibinfo{author}{\bibfnamefont{A.}~\bibnamefont{Szolnoki}}, \bibnamefont{and}
  \bibinfo{author}{\bibfnamefont{L.~M.} \bibnamefont{Flor{\'{\i}a and Y.
  Moreno}}}, \bibinfo{journal}{J. R. Soc. Interface}
  \textbf{\bibinfo{volume}{10}}, \bibinfo{pages}{20120997}
  (\bibinfo{year}{2013}).

\bibitem[{\citenamefont{Nowak and May}(1992)}]{nowak_n92b}
\bibinfo{author}{\bibfnamefont{M.~A.} \bibnamefont{Nowak}} \bibnamefont{and}
  \bibinfo{author}{\bibfnamefont{R.~M.} \bibnamefont{May}},
  \bibinfo{journal}{Nature} \textbf{\bibinfo{volume}{359}},
  \bibinfo{pages}{826} (\bibinfo{year}{1992}).

\bibitem[{\citenamefont{Nowak}(2006)}]{nowak_s06}
\bibinfo{author}{\bibfnamefont{M.~A.} \bibnamefont{Nowak}},
  \bibinfo{journal}{Science} \textbf{\bibinfo{volume}{314}},
  \bibinfo{pages}{1560} (\bibinfo{year}{2006}).

\bibitem[{\citenamefont{Nowak et~al.}(1994)\citenamefont{Nowak, Bonhoeffer, and
  May}}]{nowak_ijbc94}
\bibinfo{author}{\bibfnamefont{M.~A.} \bibnamefont{Nowak}},
  \bibinfo{author}{\bibfnamefont{S.}~\bibnamefont{Bonhoeffer}},
  \bibnamefont{and} \bibinfo{author}{\bibfnamefont{R.~M.} \bibnamefont{May}},
  \bibinfo{journal}{Int. J. Bifurcat. Chaos} \textbf{\bibinfo{volume}{4}},
  \bibinfo{pages}{33} (\bibinfo{year}{1994}).

\bibitem[{\citenamefont{Szab{\'o} and T{\H{o}}ke}(1998)}]{szabo_pre98}
\bibinfo{author}{\bibfnamefont{G.}~\bibnamefont{Szab{\'o}}} \bibnamefont{and}
  \bibinfo{author}{\bibfnamefont{C.}~\bibnamefont{T{\H{o}}ke}},
  \bibinfo{journal}{Phys. Rev. E} \textbf{\bibinfo{volume}{58}},
  \bibinfo{pages}{69} (\bibinfo{year}{1998}).

\bibitem[{\citenamefont{Santos and Pacheco}(2005)}]{santos_prl05}
\bibinfo{author}{\bibfnamefont{F.~C.} \bibnamefont{Santos}} \bibnamefont{and}
  \bibinfo{author}{\bibfnamefont{J.~M.} \bibnamefont{Pacheco}},
  \bibinfo{journal}{Phys. Rev. Lett.} \textbf{\bibinfo{volume}{95}},
  \bibinfo{pages}{098104} (\bibinfo{year}{2005}).

\bibitem[{\citenamefont{Santos et~al.}(2006{\natexlab{a}})\citenamefont{Santos,
  Pacheco, and Lenaerts}}]{santos_pnas06}
\bibinfo{author}{\bibfnamefont{F.~C.} \bibnamefont{Santos}},
  \bibinfo{author}{\bibfnamefont{J.~M.} \bibnamefont{Pacheco}},
  \bibnamefont{and} \bibinfo{author}{\bibfnamefont{T.}~\bibnamefont{Lenaerts}},
  \bibinfo{journal}{Proc. Natl. Acad. Sci. USA} \textbf{\bibinfo{volume}{103}},
  \bibinfo{pages}{3490} (\bibinfo{year}{2006}{\natexlab{a}}).

\bibitem[{\citenamefont{Castellano et~al.}(2009)\citenamefont{Castellano,
  Fortunato, and Loreto}}]{castellano_rmp09}
\bibinfo{author}{\bibfnamefont{C.}~\bibnamefont{Castellano}},
  \bibinfo{author}{\bibfnamefont{S.}~\bibnamefont{Fortunato}},
  \bibnamefont{and} \bibinfo{author}{\bibfnamefont{V.}~\bibnamefont{Loreto}},
  \bibinfo{journal}{Rev. Mod. Phys.} \textbf{\bibinfo{volume}{81}},
  \bibinfo{pages}{591} (\bibinfo{year}{2009}).

\bibitem[{\citenamefont{Perc and Grigolini}(2013)}]{perc_csf13}
\bibinfo{author}{\bibfnamefont{M.}~\bibnamefont{Perc}} \bibnamefont{and}
  \bibinfo{author}{\bibfnamefont{P.}~\bibnamefont{Grigolini}},
  \bibinfo{journal}{Chaos, Solitons \& Fractals} \textbf{\bibinfo{volume}{56}},
  \bibinfo{pages}{1} (\bibinfo{year}{2013}).

\bibitem[{\citenamefont{Abramson and Kuperman}(2001)}]{abramson_pre01}
\bibinfo{author}{\bibfnamefont{G.}~\bibnamefont{Abramson}} \bibnamefont{and}
  \bibinfo{author}{\bibfnamefont{M.}~\bibnamefont{Kuperman}},
  \bibinfo{journal}{Phys. Rev. E} \textbf{\bibinfo{volume}{63}},
  \bibinfo{pages}{030901(R)} (\bibinfo{year}{2001}).

\bibitem[{\citenamefont{Kim et~al.}(2002)\citenamefont{Kim, Trusina, Holme,
  Minnhagen, Chung, and Choi}}]{kim_bj_pre02}
\bibinfo{author}{\bibfnamefont{B.~J.} \bibnamefont{Kim}},
  \bibinfo{author}{\bibfnamefont{A.}~\bibnamefont{Trusina}},
  \bibinfo{author}{\bibfnamefont{P.}~\bibnamefont{Holme}},
  \bibinfo{author}{\bibfnamefont{P.}~\bibnamefont{Minnhagen}},
  \bibinfo{author}{\bibfnamefont{J.~S.} \bibnamefont{Chung}}, \bibnamefont{and}
  \bibinfo{author}{\bibfnamefont{M.~Y.} \bibnamefont{Choi}},
  \bibinfo{journal}{Phys. Rev. E} \textbf{\bibinfo{volume}{66}},
  \bibinfo{pages}{021907} (\bibinfo{year}{2002}).

\bibitem[{\citenamefont{Masuda and Aihara}(2003)}]{masuda_pla03}
\bibinfo{author}{\bibfnamefont{N.}~\bibnamefont{Masuda}} \bibnamefont{and}
  \bibinfo{author}{\bibfnamefont{K.}~\bibnamefont{Aihara}},
  \bibinfo{journal}{Phys. Lett. A} \textbf{\bibinfo{volume}{313}},
  \bibinfo{pages}{55} (\bibinfo{year}{2003}).

\bibitem[{\citenamefont{Tomassini et~al.}(2006)\citenamefont{Tomassini, Luthi,
  and Giacobini}}]{tomassini_pre06}
\bibinfo{author}{\bibfnamefont{M.}~\bibnamefont{Tomassini}},
  \bibinfo{author}{\bibfnamefont{L.}~\bibnamefont{Luthi}}, \bibnamefont{and}
  \bibinfo{author}{\bibfnamefont{M.}~\bibnamefont{Giacobini}},
  \bibinfo{journal}{Phys. Rev. E} \textbf{\bibinfo{volume}{73}},
  \bibinfo{pages}{016132} (\bibinfo{year}{2006}).

\bibitem[{\citenamefont{Vukov et~al.}(2006)\citenamefont{Vukov, Szab{\'o}, and
  Szolnoki}}]{vukov_pre06}
\bibinfo{author}{\bibfnamefont{J.}~\bibnamefont{Vukov}},
  \bibinfo{author}{\bibfnamefont{G.}~\bibnamefont{Szab{\'o}}},
  \bibnamefont{and} \bibinfo{author}{\bibfnamefont{A.}~\bibnamefont{Szolnoki}},
  \bibinfo{journal}{Phys. Rev. E} \textbf{\bibinfo{volume}{73}},
  \bibinfo{pages}{067103} (\bibinfo{year}{2006}).

\bibitem[{\citenamefont{Fu et~al.}(2007{\natexlab{a}})\citenamefont{Fu, Liu,
  and Wang}}]{fu_epjb07}
\bibinfo{author}{\bibfnamefont{F.}~\bibnamefont{Fu}},
  \bibinfo{author}{\bibfnamefont{L.-H.} \bibnamefont{Liu}}, \bibnamefont{and}
  \bibinfo{author}{\bibfnamefont{L.}~\bibnamefont{Wang}},
  \bibinfo{journal}{Eur. Phys. J. B} \textbf{\bibinfo{volume}{56}},
  \bibinfo{pages}{367} (\bibinfo{year}{2007}{\natexlab{a}}).

\bibitem[{\citenamefont{Vukov et~al.}(2008)\citenamefont{Vukov, Szab{\'o}, and
  Szolnoki}}]{vukov_pre08}
\bibinfo{author}{\bibfnamefont{J.}~\bibnamefont{Vukov}},
  \bibinfo{author}{\bibfnamefont{G.}~\bibnamefont{Szab{\'o}}},
  \bibnamefont{and} \bibinfo{author}{\bibfnamefont{A.}~\bibnamefont{Szolnoki}},
  \bibinfo{journal}{Phys. Rev. E} \textbf{\bibinfo{volume}{77}},
  \bibinfo{pages}{026109} (\bibinfo{year}{2008}).

\bibitem[{\citenamefont{G{\'o}mez-Garde{\~n}es
  et~al.}(2007)\citenamefont{G{\'o}mez-Garde{\~n}es, Campillo, Flor{\' \i}a,
  and Moreno}}]{gomez-gardenes_prl07}
\bibinfo{author}{\bibfnamefont{J.}~\bibnamefont{G{\'o}mez-Garde{\~n}es}},
  \bibinfo{author}{\bibfnamefont{M.}~\bibnamefont{Campillo}},
  \bibinfo{author}{\bibfnamefont{L.~M.} \bibnamefont{Flor{\' \i}a}},
  \bibnamefont{and} \bibinfo{author}{\bibfnamefont{Y.}~\bibnamefont{Moreno}},
  \bibinfo{journal}{Phys. Rev. Lett.} \textbf{\bibinfo{volume}{98}},
  \bibinfo{pages}{108103} (\bibinfo{year}{2007}).

\bibitem[{\citenamefont{Rong et~al.}(2007)\citenamefont{Rong, Li, and
  Wang}}]{rong_pre07}
\bibinfo{author}{\bibfnamefont{Z.}~\bibnamefont{Rong}},
  \bibinfo{author}{\bibfnamefont{X.}~\bibnamefont{Li}}, \bibnamefont{and}
  \bibinfo{author}{\bibfnamefont{X.}~\bibnamefont{Wang}},
  \bibinfo{journal}{Phys. Rev. E} \textbf{\bibinfo{volume}{76}},
  \bibinfo{pages}{027101} (\bibinfo{year}{2007}).

\bibitem[{\citenamefont{Masuda}(2007)}]{masuda_prsb07}
\bibinfo{author}{\bibfnamefont{N.}~\bibnamefont{Masuda}},
  \bibinfo{journal}{Proc. R. Soc. B} \textbf{\bibinfo{volume}{274}},
  \bibinfo{pages}{1815} (\bibinfo{year}{2007}).

\bibitem[{\citenamefont{Tomassini et~al.}(2007)\citenamefont{Tomassini, Luthi,
  and Pestelacci}}]{tomassini_ijmpc07}
\bibinfo{author}{\bibfnamefont{M.}~\bibnamefont{Tomassini}},
  \bibinfo{author}{\bibfnamefont{L.}~\bibnamefont{Luthi}}, \bibnamefont{and}
  \bibinfo{author}{\bibfnamefont{E.}~\bibnamefont{Pestelacci}},
  \bibinfo{journal}{Int. J. Mod. Phys. C} \textbf{\bibinfo{volume}{18}},
  \bibinfo{pages}{1173} (\bibinfo{year}{2007}).

\bibitem[{\citenamefont{Szolnoki et~al.}(2008)\citenamefont{Szolnoki, Perc, and
  Danku}}]{szolnoki_pa08}
\bibinfo{author}{\bibfnamefont{A.}~\bibnamefont{Szolnoki}},
  \bibinfo{author}{\bibfnamefont{M.}~\bibnamefont{Perc}}, \bibnamefont{and}
  \bibinfo{author}{\bibfnamefont{Z.}~\bibnamefont{Danku}},
  \bibinfo{journal}{Physica A} \textbf{\bibinfo{volume}{387}},
  \bibinfo{pages}{2075} (\bibinfo{year}{2008}).

\bibitem[{\citenamefont{Assenza et~al.}(2008)\citenamefont{Assenza,
  G{\'o}mez-Garde{\~n}es, and Latora}}]{assenza_pre08}
\bibinfo{author}{\bibfnamefont{S.}~\bibnamefont{Assenza}},
  \bibinfo{author}{\bibfnamefont{J.}~\bibnamefont{G{\'o}mez-Garde{\~n}es}},
  \bibnamefont{and} \bibinfo{author}{\bibfnamefont{V.}~\bibnamefont{Latora}},
  \bibinfo{journal}{Phys. Rev. E} \textbf{\bibinfo{volume}{78}},
  \bibinfo{pages}{017101} (\bibinfo{year}{2008}).

\bibitem[{\citenamefont{Santos et~al.}(2008)\citenamefont{Santos, Santos, and
  Pacheco}}]{santos_n08}
\bibinfo{author}{\bibfnamefont{F.~C.} \bibnamefont{Santos}},
  \bibinfo{author}{\bibfnamefont{M.~D.} \bibnamefont{Santos}},
  \bibnamefont{and} \bibinfo{author}{\bibfnamefont{J.~M.}
  \bibnamefont{Pacheco}}, \bibinfo{journal}{Nature}
  \textbf{\bibinfo{volume}{454}}, \bibinfo{pages}{213} (\bibinfo{year}{2008}).

\bibitem[{\citenamefont{Pe{\~n}a et~al.}(2009)\citenamefont{Pe{\~n}a, Volken,
  Pestelacci, and Tomassini}}]{pena_pre09}
\bibinfo{author}{\bibfnamefont{J.}~\bibnamefont{Pe{\~n}a}},
  \bibinfo{author}{\bibfnamefont{H.}~\bibnamefont{Volken}},
  \bibinfo{author}{\bibfnamefont{E.}~\bibnamefont{Pestelacci}},
  \bibnamefont{and}
  \bibinfo{author}{\bibfnamefont{M.}~\bibnamefont{Tomassini}},
  \bibinfo{journal}{Phys. Rev. E} \textbf{\bibinfo{volume}{80}},
  \bibinfo{pages}{016110} (\bibinfo{year}{2009}).

\bibitem[{\citenamefont{Poncela et~al.}(2011)\citenamefont{Poncela,
  G{\'o}mez-Garde{\~n}es, and Moreno}}]{poncela_pre11}
\bibinfo{author}{\bibfnamefont{J.}~\bibnamefont{Poncela}},
  \bibinfo{author}{\bibfnamefont{J.}~\bibnamefont{G{\'o}mez-Garde{\~n}es}},
  \bibnamefont{and} \bibinfo{author}{\bibfnamefont{Y.}~\bibnamefont{Moreno}},
  \bibinfo{journal}{Phys. Rev. E} \textbf{\bibinfo{volume}{83}},
  \bibinfo{pages}{057101} (\bibinfo{year}{2011}).

\bibitem[{\citenamefont{Poncela et~al.}(2009)\citenamefont{Poncela,
  G{\'o}mez-Garde{\~n}es, Flor{\' \i}a, Moreno, and
  S{\'a}nchez}}]{poncela_epl09}
\bibinfo{author}{\bibfnamefont{J.}~\bibnamefont{Poncela}},
  \bibinfo{author}{\bibfnamefont{J.}~\bibnamefont{G{\'o}mez-Garde{\~n}es}},
  \bibinfo{author}{\bibfnamefont{L.~M.} \bibnamefont{Flor{\' \i}a}},
  \bibinfo{author}{\bibfnamefont{Y.}~\bibnamefont{Moreno}}, \bibnamefont{and}
  \bibinfo{author}{\bibfnamefont{A.}~\bibnamefont{S{\'a}nchez}},
  \bibinfo{journal}{EPL} \textbf{\bibinfo{volume}{88}}, \bibinfo{pages}{38003}
  (\bibinfo{year}{2009}).

\bibitem[{\citenamefont{Brede}(2011)}]{brede_epl11}
\bibinfo{author}{\bibfnamefont{M.}~\bibnamefont{Brede}}, \bibinfo{journal}{EPL}
  \textbf{\bibinfo{volume}{94}}, \bibinfo{pages}{30003} (\bibinfo{year}{2011}).

\bibitem[{\citenamefont{Tanimoto et~al.}(2012)\citenamefont{Tanimoto, Brede,
  and Yamauchi}}]{tanimoto_pre12}
\bibinfo{author}{\bibfnamefont{J.}~\bibnamefont{Tanimoto}},
  \bibinfo{author}{\bibfnamefont{M.}~\bibnamefont{Brede}}, \bibnamefont{and}
  \bibinfo{author}{\bibfnamefont{A.}~\bibnamefont{Yamauchi}},
  \bibinfo{journal}{Phys. Rev. E} \textbf{\bibinfo{volume}{85}},
  \bibinfo{pages}{032101} (\bibinfo{year}{2012}).

\bibitem[{\citenamefont{Pinheiro et~al.}(2012)\citenamefont{Pinheiro, Pacheco,
  and Santos}}]{pinheiro_pone12}
\bibinfo{author}{\bibfnamefont{F.}~\bibnamefont{Pinheiro}},
  \bibinfo{author}{\bibfnamefont{J.}~\bibnamefont{Pacheco}}, \bibnamefont{and}
  \bibinfo{author}{\bibfnamefont{F.}~\bibnamefont{Santos}},
  \bibinfo{journal}{PLoS ONE} \textbf{\bibinfo{volume}{7}},
  \bibinfo{pages}{e32114} (\bibinfo{year}{2012}).

\bibitem[{\citenamefont{Simko and Csermely}(2013)}]{simko_pone13}
\bibinfo{author}{\bibfnamefont{G.~I.} \bibnamefont{Simko}} \bibnamefont{and}
  \bibinfo{author}{\bibfnamefont{P.}~\bibnamefont{Csermely}},
  \bibinfo{journal}{PLoS ONE} \textbf{\bibinfo{volume}{8}},
  \bibinfo{pages}{e67159} (\bibinfo{year}{2013}).

\bibitem[{\citenamefont{Ebel and Bornholdt}(2002)}]{ebel_pre02}
\bibinfo{author}{\bibfnamefont{H.}~\bibnamefont{Ebel}} \bibnamefont{and}
  \bibinfo{author}{\bibfnamefont{S.}~\bibnamefont{Bornholdt}},
  \bibinfo{journal}{Phys. Rev. E} \textbf{\bibinfo{volume}{66}},
  \bibinfo{pages}{056118} (\bibinfo{year}{2002}).

\bibitem[{\citenamefont{Zimmermann et~al.}(2004)\citenamefont{Zimmermann,
  Egu{\'{\i}}luz, and Miguel}}]{zimmermann_pre04}
\bibinfo{author}{\bibfnamefont{M.~G.} \bibnamefont{Zimmermann}},
  \bibinfo{author}{\bibfnamefont{V.}~\bibnamefont{Egu{\'{\i}}luz}},
  \bibnamefont{and} \bibinfo{author}{\bibfnamefont{M.~S.}
  \bibnamefont{Miguel}}, \bibinfo{journal}{Phys. Rev. E}
  \textbf{\bibinfo{volume}{69}}, \bibinfo{pages}{065102(R)}
  (\bibinfo{year}{2004}).

\bibitem[{\citenamefont{Szab{\'o} and Vukov}(2004)}]{szabo_pre04b}
\bibinfo{author}{\bibfnamefont{G.}~\bibnamefont{Szab{\'o}}} \bibnamefont{and}
  \bibinfo{author}{\bibfnamefont{J.}~\bibnamefont{Vukov}},
  \bibinfo{journal}{Phys. Rev. E} \textbf{\bibinfo{volume}{69}},
  \bibinfo{pages}{036107} (\bibinfo{year}{2004}).

\bibitem[{\citenamefont{Pacheco et~al.}(2006)\citenamefont{Pacheco, Traulsen,
  and Nowak}}]{pacheco_prl06}
\bibinfo{author}{\bibfnamefont{J.~M.} \bibnamefont{Pacheco}},
  \bibinfo{author}{\bibfnamefont{A.}~\bibnamefont{Traulsen}}, \bibnamefont{and}
  \bibinfo{author}{\bibfnamefont{M.~A.} \bibnamefont{Nowak}},
  \bibinfo{journal}{Phys. Rev. Lett.} \textbf{\bibinfo{volume}{97}},
  \bibinfo{pages}{258103} (\bibinfo{year}{2006}).

\bibitem[{\citenamefont{Santos et~al.}(2006{\natexlab{b}})\citenamefont{Santos,
  Pacheco, and Lenaerts}}]{santos_ploscb06}
\bibinfo{author}{\bibfnamefont{F.~C.} \bibnamefont{Santos}},
  \bibinfo{author}{\bibfnamefont{J.~M.} \bibnamefont{Pacheco}},
  \bibnamefont{and} \bibinfo{author}{\bibfnamefont{T.}~\bibnamefont{Lenaerts}},
  \bibinfo{journal}{PLoS Comput. Biol.} \textbf{\bibinfo{volume}{2}},
  \bibinfo{pages}{1284} (\bibinfo{year}{2006}{\natexlab{b}}).

\bibitem[{\citenamefont{Fu et~al.}(2007{\natexlab{b}})\citenamefont{Fu, Chen,
  Liu, and Wang}}]{fu_pa07}
\bibinfo{author}{\bibfnamefont{F.}~\bibnamefont{Fu}},
  \bibinfo{author}{\bibfnamefont{X.}~\bibnamefont{Chen}},
  \bibinfo{author}{\bibfnamefont{L.}~\bibnamefont{Liu}}, \bibnamefont{and}
  \bibinfo{author}{\bibfnamefont{L.}~\bibnamefont{Wang}},
  \bibinfo{journal}{Physica A} \textbf{\bibinfo{volume}{383}},
  \bibinfo{pages}{651} (\bibinfo{year}{2007}{\natexlab{b}}).

\bibitem[{\citenamefont{Tanimoto}(2007)}]{tanimoto_pre07}
\bibinfo{author}{\bibfnamefont{J.}~\bibnamefont{Tanimoto}},
  \bibinfo{journal}{Phys. Rev. E} \textbf{\bibinfo{volume}{76}},
  \bibinfo{pages}{021126} (\bibinfo{year}{2007}).

\bibitem[{\citenamefont{Fu et~al.}(2008)\citenamefont{Fu, Hauert, Nowak, and
  Wang}}]{fu_pre08b}
\bibinfo{author}{\bibfnamefont{F.}~\bibnamefont{Fu}},
  \bibinfo{author}{\bibfnamefont{C.}~\bibnamefont{Hauert}},
  \bibinfo{author}{\bibfnamefont{M.~A.} \bibnamefont{Nowak}}, \bibnamefont{and}
  \bibinfo{author}{\bibfnamefont{L.}~\bibnamefont{Wang}},
  \bibinfo{journal}{Phys. Rev. E} \textbf{\bibinfo{volume}{78}},
  \bibinfo{pages}{026117} (\bibinfo{year}{2008}).

\bibitem[{\citenamefont{Fu et~al.}(2009)\citenamefont{Fu, Wu, and
  Wang}}]{fu_pre09}
\bibinfo{author}{\bibfnamefont{F.}~\bibnamefont{Fu}},
  \bibinfo{author}{\bibfnamefont{T.}~\bibnamefont{Wu}}, \bibnamefont{and}
  \bibinfo{author}{\bibfnamefont{L.}~\bibnamefont{Wang}},
  \bibinfo{journal}{Phys. Rev. E} \textbf{\bibinfo{volume}{79}},
  \bibinfo{pages}{036101} (\bibinfo{year}{2009}).

\bibitem[{\citenamefont{Vukov and Szab{\'o}}(2005)}]{vukov_pre05}
\bibinfo{author}{\bibfnamefont{J.}~\bibnamefont{Vukov}} \bibnamefont{and}
  \bibinfo{author}{\bibfnamefont{G.}~\bibnamefont{Szab{\'o}}},
  \bibinfo{journal}{Phys. Rev. E} \textbf{\bibinfo{volume}{71}},
  \bibinfo{pages}{036133} (\bibinfo{year}{2005}).

\bibitem[{\citenamefont{Lee et~al.}(2011)\citenamefont{Lee, Holme, and
  Wu}}]{lee_s_prl11}
\bibinfo{author}{\bibfnamefont{S.}~\bibnamefont{Lee}},
  \bibinfo{author}{\bibfnamefont{P.}~\bibnamefont{Holme}}, \bibnamefont{and}
  \bibinfo{author}{\bibfnamefont{Z.-X.} \bibnamefont{Wu}},
  \bibinfo{journal}{Phys. Rev. Lett.} \textbf{\bibinfo{volume}{106}},
  \bibinfo{pages}{028702} (\bibinfo{year}{2011}).

\bibitem[{\citenamefont{G{\'o}mez-Garde{\~n}es
  et~al.}(2011{\natexlab{a}})\citenamefont{G{\'o}mez-Garde{\~n}es, Romance,
  Criado, Vilone, and S{\'a}nchez}}]{gomez-gardenes_c11}
\bibinfo{author}{\bibfnamefont{J.}~\bibnamefont{G{\'o}mez-Garde{\~n}es}},
  \bibinfo{author}{\bibfnamefont{M.}~\bibnamefont{Romance}},
  \bibinfo{author}{\bibfnamefont{R.}~\bibnamefont{Criado}},
  \bibinfo{author}{\bibfnamefont{D.}~\bibnamefont{Vilone}}, \bibnamefont{and}
  \bibinfo{author}{\bibfnamefont{A.}~\bibnamefont{S{\'a}nchez}},
  \bibinfo{journal}{Chaos} \textbf{\bibinfo{volume}{21}},
  \bibinfo{pages}{016113} (\bibinfo{year}{2011}{\natexlab{a}}).

\bibitem[{\citenamefont{G{\'o}mez-Garde{\~n}es
  et~al.}(2011{\natexlab{b}})\citenamefont{G{\'o}mez-Garde{\~n}es, Vilone, and
  S{\'a}nchez}}]{gomez-gardenes_epl11}
\bibinfo{author}{\bibfnamefont{J.}~\bibnamefont{G{\'o}mez-Garde{\~n}es}},
  \bibinfo{author}{\bibfnamefont{D.}~\bibnamefont{Vilone}}, \bibnamefont{and}
  \bibinfo{author}{\bibfnamefont{A.}~\bibnamefont{S{\'a}nchez}},
  \bibinfo{journal}{EPL} \textbf{\bibinfo{volume}{95}}, \bibinfo{pages}{68003}
  (\bibinfo{year}{2011}{\natexlab{b}}).

\bibitem[{\citenamefont{Gracia-L{\'a}zaro
  et~al.}(2012{\natexlab{a}})\citenamefont{Gracia-L{\'a}zaro, Cuesta,
  S{\'a}nchez, and Moreno}}]{gracia-lazaro_srep12}
\bibinfo{author}{\bibfnamefont{C.}~\bibnamefont{Gracia-L{\'a}zaro}},
  \bibinfo{author}{\bibfnamefont{J.}~\bibnamefont{Cuesta}},
  \bibinfo{author}{\bibfnamefont{A.}~\bibnamefont{S{\'a}nchez}},
  \bibnamefont{and} \bibinfo{author}{\bibfnamefont{Y.}~\bibnamefont{Moreno}},
  \bibinfo{journal}{Sci. Rep.} \textbf{\bibinfo{volume}{2}},
  \bibinfo{pages}{325} (\bibinfo{year}{2012}{\natexlab{a}}).

\bibitem[{\citenamefont{Gracia-L{\'a}zaro
  et~al.}(2012{\natexlab{b}})\citenamefont{Gracia-L{\'a}zaro, Ferrer, Ruiz,
  Taranc{\'o}n, Cuesta, S{\'a}nchez, and Moreno}}]{gracia-lazaro_pnas12}
\bibinfo{author}{\bibfnamefont{C.}~\bibnamefont{Gracia-L{\'a}zaro}},
  \bibinfo{author}{\bibfnamefont{A.}~\bibnamefont{Ferrer}},
  \bibinfo{author}{\bibfnamefont{G.}~\bibnamefont{Ruiz}},
  \bibinfo{author}{\bibfnamefont{A.}~\bibnamefont{Taranc{\'o}n}},
  \bibinfo{author}{\bibfnamefont{J.}~\bibnamefont{Cuesta}},
  \bibinfo{author}{\bibfnamefont{A.}~\bibnamefont{S{\'a}nchez}},
  \bibnamefont{and} \bibinfo{author}{\bibfnamefont{Y.}~\bibnamefont{Moreno}},
  \bibinfo{journal}{Proc. Natl. Acad. Sci. USA} \textbf{\bibinfo{volume}{109}},
  \bibinfo{pages}{12922} (\bibinfo{year}{2012}{\natexlab{b}}).

\bibitem[{\citenamefont{Wang et~al.}(2012{\natexlab{a}})\citenamefont{Wang,
  Szolnoki, and Perc}}]{wang_z_epl12}
\bibinfo{author}{\bibfnamefont{Z.}~\bibnamefont{Wang}},
  \bibinfo{author}{\bibfnamefont{A.}~\bibnamefont{Szolnoki}}, \bibnamefont{and}
  \bibinfo{author}{\bibfnamefont{M.}~\bibnamefont{Perc}},
  \bibinfo{journal}{EPL} \textbf{\bibinfo{volume}{97}}, \bibinfo{pages}{48001}
  (\bibinfo{year}{2012}{\natexlab{a}}).

\bibitem[{\citenamefont{G{\'o}mez-Garde{\~n}es
  et~al.}(2012{\natexlab{a}})\citenamefont{G{\'o}mez-Garde{\~n}es, Reinares,
  Arenas, and Flor{\' \i}a}}]{gomez-gardenes_srep12}
\bibinfo{author}{\bibfnamefont{J.}~\bibnamefont{G{\'o}mez-Garde{\~n}es}},
  \bibinfo{author}{\bibfnamefont{I.}~\bibnamefont{Reinares}},
  \bibinfo{author}{\bibfnamefont{A.}~\bibnamefont{Arenas}}, \bibnamefont{and}
  \bibinfo{author}{\bibfnamefont{L.~M.} \bibnamefont{Flor{\' \i}a}},
  \bibinfo{journal}{Sci. Rep.} \textbf{\bibinfo{volume}{2}},
  \bibinfo{pages}{620} (\bibinfo{year}{2012}{\natexlab{a}}).

\bibitem[{\citenamefont{G{\'o}mez-Garde{\~n}es
  et~al.}(2012{\natexlab{b}})\citenamefont{G{\'o}mez-Garde{\~n}es,
  Gracia-L{\'a}zaro, Flor{\' \i}a, and Moreno}}]{gomez-gardenes_pre12}
\bibinfo{author}{\bibfnamefont{J.}~\bibnamefont{G{\'o}mez-Garde{\~n}es}},
  \bibinfo{author}{\bibfnamefont{C.}~\bibnamefont{Gracia-L{\'a}zaro}},
  \bibinfo{author}{\bibfnamefont{L.~M.} \bibnamefont{Flor{\' \i}a}},
  \bibnamefont{and} \bibinfo{author}{\bibfnamefont{Y.}~\bibnamefont{Moreno}},
  \bibinfo{journal}{Phys. Rev. E} \textbf{\bibinfo{volume}{86}},
  \bibinfo{pages}{056113} (\bibinfo{year}{2012}{\natexlab{b}}).

\bibitem[{\citenamefont{Wang et~al.}(2012{\natexlab{b}})\citenamefont{Wang,
  Chen, and Wang}}]{wang_b_jsm12}
\bibinfo{author}{\bibfnamefont{B.}~\bibnamefont{Wang}},
  \bibinfo{author}{\bibfnamefont{X.}~\bibnamefont{Chen}}, \bibnamefont{and}
  \bibinfo{author}{\bibfnamefont{L.}~\bibnamefont{Wang}}, \bibinfo{journal}{J.
  Stat. Mech.} \textbf{\bibinfo{volume}{2012}}, \bibinfo{pages}{P11017}
  (\bibinfo{year}{2012}{\natexlab{b}}).

\bibitem[{\citenamefont{Wang et~al.}(2013{\natexlab{a}})\citenamefont{Wang,
  Szolnoki, and Perc}}]{wang_z_srep13}
\bibinfo{author}{\bibfnamefont{Z.}~\bibnamefont{Wang}},
  \bibinfo{author}{\bibfnamefont{A.}~\bibnamefont{Szolnoki}}, \bibnamefont{and}
  \bibinfo{author}{\bibfnamefont{M.}~\bibnamefont{Perc}},
  \bibinfo{journal}{Sci. Rep.} \textbf{\bibinfo{volume}{3}},
  \bibinfo{pages}{1183} (\bibinfo{year}{2013}{\natexlab{a}}).

\bibitem[{\citenamefont{Wang et~al.}(2013{\natexlab{b}})\citenamefont{Wang,
  Szolnoki, and Perc}}]{wang_z_srep13b}
\bibinfo{author}{\bibfnamefont{Z.}~\bibnamefont{Wang}},
  \bibinfo{author}{\bibfnamefont{A.}~\bibnamefont{Szolnoki}}, \bibnamefont{and}
  \bibinfo{author}{\bibfnamefont{M.}~\bibnamefont{Perc}},
  \bibinfo{journal}{Sci. Rep.} \textbf{\bibinfo{volume}{3}},
  \bibinfo{pages}{2470} (\bibinfo{year}{2013}{\natexlab{b}}).

\bibitem[{\citenamefont{Jiang and Perc}(2013)}]{jiang_ll_srep13}
\bibinfo{author}{\bibfnamefont{L.-L.} \bibnamefont{Jiang}} \bibnamefont{and}
  \bibinfo{author}{\bibfnamefont{M.}~\bibnamefont{Perc}},
  \bibinfo{journal}{Sci. Rep.} \textbf{\bibinfo{volume}{3}},
  \bibinfo{pages}{2483} (\bibinfo{year}{2013}).

\bibitem[{\citenamefont{Szolnoki and Perc}(2013)}]{szolnoki_njp13}
\bibinfo{author}{\bibfnamefont{A.}~\bibnamefont{Szolnoki}} \bibnamefont{and}
  \bibinfo{author}{\bibfnamefont{M.}~\bibnamefont{Perc}}, \bibinfo{journal}{New
  J. Phys.} \textbf{\bibinfo{volume}{15}}, \bibinfo{pages}{053010}
  (\bibinfo{year}{2013}).

\bibitem[{\citenamefont{Buldyrev et~al.}(2010)\citenamefont{Buldyrev, Parshani,
  Paul, Stanley, and Havlin}}]{buldyrev_n10}
\bibinfo{author}{\bibfnamefont{S.~V.} \bibnamefont{Buldyrev}},
  \bibinfo{author}{\bibfnamefont{R.}~\bibnamefont{Parshani}},
  \bibinfo{author}{\bibfnamefont{G.}~\bibnamefont{Paul}},
  \bibinfo{author}{\bibfnamefont{H.~E.} \bibnamefont{Stanley}},
  \bibnamefont{and} \bibinfo{author}{\bibfnamefont{S.}~\bibnamefont{Havlin}},
  \bibinfo{journal}{Nature} \textbf{\bibinfo{volume}{464}},
  \bibinfo{pages}{1025} (\bibinfo{year}{2010}).

\bibitem[{\citenamefont{Li et~al.}(2012)\citenamefont{Li, Bashan, and
  Buldyrev}}]{li_w_prl12}
\bibinfo{author}{\bibfnamefont{W.}~\bibnamefont{Li}},
  \bibinfo{author}{\bibfnamefont{A.}~\bibnamefont{Bashan}}, \bibnamefont{and}
  \bibinfo{author}{\bibfnamefont{S.~V.} \bibnamefont{Buldyrev}},
  \bibinfo{journal}{Phys. Rev. Lett.} \textbf{\bibinfo{volume}{108}},
  \bibinfo{pages}{228702} (\bibinfo{year}{2012}).

\bibitem[{\citenamefont{Parshani et~al.}(2011)\citenamefont{Parshani, Buldyrev,
  and Havlin}}]{parshani_pnas11}
\bibinfo{author}{\bibfnamefont{R.}~\bibnamefont{Parshani}},
  \bibinfo{author}{\bibfnamefont{S.~V.} \bibnamefont{Buldyrev}},
  \bibnamefont{and} \bibinfo{author}{\bibfnamefont{S.}~\bibnamefont{Havlin}},
  \bibinfo{journal}{Proc. Natl. Acad. Sci. USA} \textbf{\bibinfo{volume}{108}},
  \bibinfo{pages}{1007} (\bibinfo{year}{2011}).

\bibitem[{\citenamefont{Brummitt et~al.}(2012)\citenamefont{Brummitt,
  \protect{D'Souza}, and Leicht}}]{brummitt_pnas12}
\bibinfo{author}{\bibfnamefont{C.~D.} \bibnamefont{Brummitt}},
  \bibinfo{author}{\bibfnamefont{R.~M.} \bibnamefont{\protect{D'Souza}}},
  \bibnamefont{and} \bibinfo{author}{\bibfnamefont{E.~A.}
  \bibnamefont{Leicht}}, \bibinfo{journal}{Proc. Natl. Acad. Sci. USA}
  \textbf{\bibinfo{volume}{109}}, \bibinfo{pages}{E680} (\bibinfo{year}{2012}).

\bibitem[{\citenamefont{Parshani et~al.}(2010)\citenamefont{Parshani, Buldyrev,
  and Havlin}}]{parshani_prl10}
\bibinfo{author}{\bibfnamefont{R.}~\bibnamefont{Parshani}},
  \bibinfo{author}{\bibfnamefont{S.~V.} \bibnamefont{Buldyrev}},
  \bibnamefont{and} \bibinfo{author}{\bibfnamefont{S.}~\bibnamefont{Havlin}},
  \bibinfo{journal}{Phys. Rev. Lett.} \textbf{\bibinfo{volume}{105}},
  \bibinfo{pages}{048701} (\bibinfo{year}{2010}).

\bibitem[{\citenamefont{Nagler et~al.}(2011)\citenamefont{Nagler, Levina, and
  Timme}}]{nagler_np11}
\bibinfo{author}{\bibfnamefont{J.}~\bibnamefont{Nagler}},
  \bibinfo{author}{\bibfnamefont{A.}~\bibnamefont{Levina}}, \bibnamefont{and}
  \bibinfo{author}{\bibfnamefont{M.}~\bibnamefont{Timme}},
  \bibinfo{journal}{Nature Physics} \textbf{\bibinfo{volume}{7}},
  \bibinfo{pages}{265} (\bibinfo{year}{2011}).

\bibitem[{\citenamefont{Cellai et~al.}(2013)\citenamefont{Cellai, L{\'o}pez,
  Zhou, Gleeson, and Bianconi}}]{cellai_pre13}
\bibinfo{author}{\bibfnamefont{D.}~\bibnamefont{Cellai}},
  \bibinfo{author}{\bibfnamefont{E.}~\bibnamefont{L{\'o}pez}},
  \bibinfo{author}{\bibfnamefont{J.}~\bibnamefont{Zhou}},
  \bibinfo{author}{\bibfnamefont{J.~P.} \bibnamefont{Gleeson}},
  \bibnamefont{and} \bibinfo{author}{\bibfnamefont{G.}~\bibnamefont{Bianconi}},
  \bibinfo{journal}{Phys. Rev. E} \textbf{\bibinfo{volume}{88}},
  \bibinfo{pages}{052811} (\bibinfo{year}{2013}).

\bibitem[{\citenamefont{Morris and Barthelemy}(2012)}]{morris_prl12}
\bibinfo{author}{\bibfnamefont{R.~G.} \bibnamefont{Morris}} \bibnamefont{and}
  \bibinfo{author}{\bibfnamefont{M.}~\bibnamefont{Barthelemy}},
  \bibinfo{journal}{Phys. Rev. Lett.} \textbf{\bibinfo{volume}{109}},
  \bibinfo{pages}{128703} (\bibinfo{year}{2012}).

\bibitem[{\citenamefont{G{\'o}mez et~al.}(2013)\citenamefont{G{\'o}mez,
  D{\'i}az-Guilera, G{\'o}mez-Garde{\~n}es, P{\'e}rez-Vicente, Moreno, and
  Arenas}}]{gomez_prl13}
\bibinfo{author}{\bibfnamefont{S.}~\bibnamefont{G{\'o}mez}},
  \bibinfo{author}{\bibfnamefont{A.}~\bibnamefont{D{\'i}az-Guilera}},
  \bibinfo{author}{\bibfnamefont{J.}~\bibnamefont{G{\'o}mez-Garde{\~n}es}},
  \bibinfo{author}{\bibfnamefont{C.}~\bibnamefont{P{\'e}rez-Vicente}},
  \bibinfo{author}{\bibfnamefont{Y.}~\bibnamefont{Moreno}}, \bibnamefont{and}
  \bibinfo{author}{\bibfnamefont{A.}~\bibnamefont{Arenas}},
  \bibinfo{journal}{Phys. Rev. Lett.} \textbf{\bibinfo{volume}{110}},
  \bibinfo{pages}{028701} (\bibinfo{year}{2013}).

\bibitem[{\citenamefont{Sun et~al.}(2011)\citenamefont{Sun, Lei, Perc, Kurths,
  and Chen}}]{sun_xj_chaos11}
\bibinfo{author}{\bibfnamefont{X.}~\bibnamefont{Sun}},
  \bibinfo{author}{\bibfnamefont{J.}~\bibnamefont{Lei}},
  \bibinfo{author}{\bibfnamefont{M.}~\bibnamefont{Perc}},
  \bibinfo{author}{\bibfnamefont{J.}~\bibnamefont{Kurths}}, \bibnamefont{and}
  \bibinfo{author}{\bibfnamefont{G.}~\bibnamefont{Chen}},
  \bibinfo{journal}{Chaos} \textbf{\bibinfo{volume}{21}},
  \bibinfo{pages}{016110} (\bibinfo{year}{2011}).

\bibitem[{\citenamefont{Granell et~al.}(2013)\citenamefont{Granell, G{\'o}mez,
  and Arenas}}]{granell_prl13}
\bibinfo{author}{\bibfnamefont{C.}~\bibnamefont{Granell}},
  \bibinfo{author}{\bibfnamefont{S.}~\bibnamefont{G{\'o}mez}},
  \bibnamefont{and} \bibinfo{author}{\bibfnamefont{A.}~\bibnamefont{Arenas}},
  \bibinfo{journal}{Phys. Rev. Lett.} \textbf{\bibinfo{volume}{111}},
  \bibinfo{pages}{128701} (\bibinfo{year}{2013}).

\bibitem[{\citenamefont{Huang et~al.}(2011)\citenamefont{Huang, Gao, Buldyrev,
  Havlin, and Stanley}}]{huang_xq_pre11}
\bibinfo{author}{\bibfnamefont{X.}~\bibnamefont{Huang}},
  \bibinfo{author}{\bibfnamefont{J.}~\bibnamefont{Gao}},
  \bibinfo{author}{\bibfnamefont{S.~V.} \bibnamefont{Buldyrev}},
  \bibinfo{author}{\bibfnamefont{S.}~\bibnamefont{Havlin}}, \bibnamefont{and}
  \bibinfo{author}{\bibfnamefont{H.~E.} \bibnamefont{Stanley}},
  \bibinfo{journal}{Phys. Rev. E} \textbf{\bibinfo{volume}{83}},
  \bibinfo{pages}{065101(R)} (\bibinfo{year}{2011}).

\bibitem[{\citenamefont{Zhou et~al.}(2012)\citenamefont{Zhou, Stanley,
  D'Agostino, and Scala}}]{zhou_d_pre12}
\bibinfo{author}{\bibfnamefont{D.}~\bibnamefont{Zhou}},
  \bibinfo{author}{\bibfnamefont{H.~E.} \bibnamefont{Stanley}},
  \bibinfo{author}{\bibfnamefont{G.}~\bibnamefont{D'Agostino}},
  \bibnamefont{and} \bibinfo{author}{\bibfnamefont{A.}~\bibnamefont{Scala}},
  \bibinfo{journal}{Phys. Rev. E} \textbf{\bibinfo{volume}{86}},
  \bibinfo{pages}{066103} (\bibinfo{year}{2012}).

\bibitem[{\citenamefont{Cozzo et~al.}(2012)\citenamefont{Cozzo, Arenas, and
  Moreno}}]{cozzo_pre13}
\bibinfo{author}{\bibfnamefont{E.}~\bibnamefont{Cozzo}},
  \bibinfo{author}{\bibfnamefont{A.}~\bibnamefont{Arenas}}, \bibnamefont{and}
  \bibinfo{author}{\bibfnamefont{Y.}~\bibnamefont{Moreno}},
  \bibinfo{journal}{Phys. Rev. E} \textbf{\bibinfo{volume}{86}},
  \bibinfo{pages}{036115} (\bibinfo{year}{2012}).

\bibitem[{\citenamefont{Nicosia et~al.}(2013)\citenamefont{Nicosia, Bianconi,
  Latora, and Barthelemy}}]{nicosia_prl13}
\bibinfo{author}{\bibfnamefont{V.}~\bibnamefont{Nicosia}},
  \bibinfo{author}{\bibfnamefont{G.}~\bibnamefont{Bianconi}},
  \bibinfo{author}{\bibfnamefont{V.}~\bibnamefont{Latora}}, \bibnamefont{and}
  \bibinfo{author}{\bibfnamefont{M.}~\bibnamefont{Barthelemy}},
  \bibinfo{journal}{Phys. Rev. Lett.} \textbf{\bibinfo{volume}{111}},
  \bibinfo{pages}{058701} (\bibinfo{year}{2013}).

\bibitem[{\citenamefont{Bianconi}(2013)}]{bianconi_pre13}
\bibinfo{author}{\bibfnamefont{G.}~\bibnamefont{Bianconi}},
  \bibinfo{journal}{Phys. Rev. E} \textbf{\bibinfo{volume}{87}},
  \bibinfo{pages}{062806} (\bibinfo{year}{2013}).

\bibitem[{\citenamefont{Radicchi and Arenas}(2013)}]{radicchi_np13}
\bibinfo{author}{\bibfnamefont{F.}~\bibnamefont{Radicchi}} \bibnamefont{and}
  \bibinfo{author}{\bibfnamefont{A.}~\bibnamefont{Arenas}},
  \bibinfo{journal}{Nature Physics} \textbf{\bibinfo{volume}{9}},
  \bibinfo{pages}{717} (\bibinfo{year}{2013}).

\bibitem[{\citenamefont{Gao et~al.}(2012)\citenamefont{Gao, Buldyrev, Stanley,
  and Havlin}}]{gao_jx_np12}
\bibinfo{author}{\bibfnamefont{J.}~\bibnamefont{Gao}},
  \bibinfo{author}{\bibfnamefont{S.~V.} \bibnamefont{Buldyrev}},
  \bibinfo{author}{\bibfnamefont{H.~E.} \bibnamefont{Stanley}},
  \bibnamefont{and} \bibinfo{author}{\bibfnamefont{S.}~\bibnamefont{Havlin}},
  \bibinfo{journal}{Nature Physics} \textbf{\bibinfo{volume}{8}},
  \bibinfo{pages}{40} (\bibinfo{year}{2012}).

\bibitem[{\citenamefont{Havlin et~al.}(2012)\citenamefont{Havlin, Kenett,
  \protect{Ben-Jacob}, Bunde, Hermann, Kurths, Kirkpatrick, Solomon, and
  Portugali}}]{havlin_pst12}
\bibinfo{author}{\bibfnamefont{S.}~\bibnamefont{Havlin}},
  \bibinfo{author}{\bibfnamefont{D.~Y.} \bibnamefont{Kenett}},
  \bibinfo{author}{\bibfnamefont{E.}~\bibnamefont{\protect{Ben-Jacob}}},
  \bibinfo{author}{\bibfnamefont{A.}~\bibnamefont{Bunde}},
  \bibinfo{author}{\bibfnamefont{H.}~\bibnamefont{Hermann}},
  \bibinfo{author}{\bibfnamefont{J.}~\bibnamefont{Kurths}},
  \bibinfo{author}{\bibfnamefont{S.}~\bibnamefont{Kirkpatrick}},
  \bibinfo{author}{\bibfnamefont{S.}~\bibnamefont{Solomon}}, \bibnamefont{and}
  \bibinfo{author}{\bibfnamefont{J.}~\bibnamefont{Portugali}},
  \bibinfo{journal}{Eur. J. Phys. Special Topics}
  \textbf{\bibinfo{volume}{214}}, \bibinfo{pages}{273} (\bibinfo{year}{2012}).

\bibitem[{\citenamefont{Helbing}(2013)}]{helbing_n13}
\bibinfo{author}{\bibfnamefont{D.}~\bibnamefont{Helbing}},
  \bibinfo{journal}{Nature} \textbf{\bibinfo{volume}{497}}, \bibinfo{pages}{51}
  (\bibinfo{year}{2013}).

\bibitem[{\citenamefont{Csermely}(2013)}]{csermely_tde13}
\bibinfo{author}{\bibfnamefont{P.}~\bibnamefont{Csermely}},
  \bibinfo{journal}{Talent Development Excellence}
  \textbf{\bibinfo{volume}{5}}, \bibinfo{pages}{115} (\bibinfo{year}{2013}).

\bibitem[{\citenamefont{Kivel{\"a} et~al.}(2013)\citenamefont{Kivel{\"a},
  Arenas, Barthelemy, Gleeson, Moreno, and Porter}}]{kivela_ax13}
\bibinfo{author}{\bibfnamefont{M.}~\bibnamefont{Kivel{\"a}}},
  \bibinfo{author}{\bibfnamefont{A.}~\bibnamefont{Arenas}},
  \bibinfo{author}{\bibfnamefont{M.}~\bibnamefont{Barthelemy}},
  \bibinfo{author}{\bibfnamefont{J.~P.} \bibnamefont{Gleeson}},
  \bibinfo{author}{\bibfnamefont{Y.}~\bibnamefont{Moreno}}, \bibnamefont{and}
  \bibinfo{author}{\bibfnamefont{M.~A.} \bibnamefont{Porter}},
  \bibinfo{journal}{Journal of Complex Networks} (\bibinfo{year}{2014}).

\bibitem[{\citenamefont{Ohtsuki
  et~al.}(2007{\natexlab{a}})\citenamefont{Ohtsuki, Nowak, and
  Pacheco}}]{ohtsuki_prl07}
\bibinfo{author}{\bibfnamefont{H.}~\bibnamefont{Ohtsuki}},
  \bibinfo{author}{\bibfnamefont{M.~A.} \bibnamefont{Nowak}}, \bibnamefont{and}
  \bibinfo{author}{\bibfnamefont{J.~M.} \bibnamefont{Pacheco}},
  \bibinfo{journal}{Phys. Rev. Lett.} \textbf{\bibinfo{volume}{98}},
  \bibinfo{pages}{108106} (\bibinfo{year}{2007}{\natexlab{a}}).

\bibitem[{\citenamefont{Ohtsuki
  et~al.}(2007{\natexlab{b}})\citenamefont{Ohtsuki, Pacheco, and
  Nowak}}]{ohtsuki_jtb07b}
\bibinfo{author}{\bibfnamefont{H.}~\bibnamefont{Ohtsuki}},
  \bibinfo{author}{\bibfnamefont{J.~M.} \bibnamefont{Pacheco}},
  \bibnamefont{and} \bibinfo{author}{\bibfnamefont{M.~A.} \bibnamefont{Nowak}},
  \bibinfo{journal}{J. Theor. Biol.} \textbf{\bibinfo{volume}{246}},
  \bibinfo{pages}{681} (\bibinfo{year}{2007}{\natexlab{b}}).
  
\bibitem[{\citenamefont{Barab{\'a}si and Albert}(1999)}]{barabasi_s99}
\bibinfo{author}{\bibfnamefont{A.-L.} \bibnamefont{Barab{\'a}si}}
  \bibnamefont{and} \bibinfo{author}{\bibfnamefont{R.}~\bibnamefont{Albert}},
  \bibinfo{journal}{Science} \textbf{\bibinfo{volume}{286}},
  \bibinfo{pages}{509} (\bibinfo{year}{1999}).

\bibitem[{\citenamefont{Xulvi-Brunet and Sokolov}(2004)}]{xulvi-brunet_pre04}
\bibinfo{author}{\bibfnamefont{R.}~\bibnamefont{Xulvi-Brunet}}
  \bibnamefont{and} \bibinfo{author}{\bibfnamefont{I.~M.}
  \bibnamefont{Sokolov}}, \bibinfo{journal}{Phys. Rev. E}
  \textbf{\bibinfo{volume}{70}}, \bibinfo{pages}{066102}
  (\bibinfo{year}{2004}).

\bibitem[{\citenamefont{Molloy and Reed}(1995)}]{molloy_rsa95}
\bibinfo{author}{\bibfnamefont{M.}~\bibnamefont{Molloy}} \bibnamefont{and}
  \bibinfo{author}{\bibfnamefont{B.}~\bibnamefont{Reed}},
  \bibinfo{journal}{Random Structures \& Algorithms}
  \textbf{\bibinfo{volume}{6}}, \bibinfo{pages}{161} (\bibinfo{year}{1995}).

\bibitem[{\citenamefont{Blume}(1993)}]{blume_l_geb93}
\bibinfo{author}{\bibfnamefont{L.~E.} \bibnamefont{Blume}},
  \bibinfo{journal}{Games Econ. Behav.} \textbf{\bibinfo{volume}{5}},
  \bibinfo{pages}{387} (\bibinfo{year}{1993}).

\bibitem[{\citenamefont{Cardillo et~al.}(2010)\citenamefont{Cardillo,
  G{\'o}mez-Garde{\~n}es, Vilone, and S{\'a}nchez}}]{cardillo_njp10}
\bibinfo{author}{\bibfnamefont{A.}~\bibnamefont{Cardillo}},
  \bibinfo{author}{\bibfnamefont{J.}~\bibnamefont{G{\'o}mez-Garde{\~n}es}},
  \bibinfo{author}{\bibfnamefont{D.}~\bibnamefont{Vilone}}, \bibnamefont{and}
  \bibinfo{author}{\bibfnamefont{A.}~\bibnamefont{S{\'a}nchez}},
  \bibinfo{journal}{New J. Phys.} \textbf{\bibinfo{volume}{12}},
  \bibinfo{pages}{103034} (\bibinfo{year}{2010}).

\bibitem[{\citenamefont{Buesser and Tomassini}(2012)}]{buesser_pre12b}
\bibinfo{author}{\bibfnamefont{P.}~\bibnamefont{Buesser}} \bibnamefont{and}
  \bibinfo{author}{\bibfnamefont{M.}~\bibnamefont{Tomassini}},
  \bibinfo{journal}{Phys. Rev. E} \textbf{\bibinfo{volume}{86}},
  \bibinfo{pages}{066107} (\bibinfo{year}{2012}).

\bibitem[{\citenamefont{Roca et~al.}(2006)\citenamefont{Roca, Cuesta, and
  S{\'a}nchez}}]{roca_prl06}
\bibinfo{author}{\bibfnamefont{C.~P.} \bibnamefont{Roca}},
  \bibinfo{author}{\bibfnamefont{J.~A.} \bibnamefont{Cuesta}},
  \bibnamefont{and}
  \bibinfo{author}{\bibfnamefont{A.}~\bibnamefont{S{\'a}nchez}},
  \bibinfo{journal}{Phys. Rev. Lett.} \textbf{\bibinfo{volume}{97}},
  \bibinfo{pages}{158701} (\bibinfo{year}{2006}).

\end{thebibliography}
\end{document}